%% file: main.tex
\documentclass[sigconf]{acmart}

\usepackage{todonotes}
\usepackage{graphicx}
\usepackage{multirow}
\usepackage{booktabs}
\usepackage{subcaption}
\usepackage{pifont}
\usepackage{siunitx}
\usepackage{array}

\newcommand{\cut}[1]{}

\newcommand{\inlineheading}[1]{\noindent\textit{\underline{#1.}} }

\setcopyright{acmlicensed}
\copyrightyear{2025}
\acmYear{20xx}
\acmDOI{XXXXXXX.XXXXXXX}
\acmConference[PACT '25]{}{November 3--6, 2025}{Irvine, CA, USA}
\acmISBN{978-1-4503-XXXX-X/20xx/xx}

\renewcommand\footnotetextcopyrightpermission[1]{} 

\begin{document}

\title{Fine-Grained Fusion: The Missing Piece in Area-Efficient State Space Model Acceleration}

\author{Robin Geens}
\authornote{Equal contribution}
\orcid{0009-0000-2450-2660}
\email{robin.geens@kuleuven.be}
\affiliation{%
  \institution{MICAS, KU Leuven}
  \city{Leuven}
  \country{Belgium}
}

\author{Arne Symons}
\authornotemark[1]
\email{arne.symons@kuleuven.be}
\orcid{0000-0002-4595-9384}
\affiliation{%
  \institution{MICAS, KU Leuven}
  \city{Leuven}
  \country{Belgium}
}

\author{Marian Verhelst}
\orcid{0000-0003-3495-9263}
\email{marian.verhelst@kuleuven.be}
\affiliation{%
  \institution{MICAS, KU Leuven}
  \city{Leuven}
  \country{Belgium}
}

\input{0-abstr.tex}




\maketitle
\pagestyle{plain} 

\input{1-intro.tex}
\input{2-background.tex}

\input{3-motivation}
\input{3.1_methodology}
\input{4-fusion}
\input{5-dse}
\input{7-concls}
\begin{acks}
This project has been partly funded by the European Research Council (ERC) under grant agreement No. 101088865, the European Union’s Horizon 2020 program under grant agreement No. 101070374, the Flanders AI Research Program, Research Foundation Flanders (FWO) under grant No. 1S37125N, and KU Leuven.
\end{acks}

\bibliographystyle{ACM-Reference-Format}
\bibliography{ref}	

\end{document}

%% file: 0-abstr.tex
\begin{abstract}
State Space Models (SSMs) offer a promising alternative to transformers for long-sequence processing. However, their efficiency remains hindered by memory-bound operations, particularly in the prefill stage. While MARCA, a recent first effort to accelerate SSMs through a dedicated hardware accelerator, achieves great speedup over high-end GPUs, an analysis into the broader accelerator design space is lacking.
This work systematically analyzes SSM acceleration opportunities both from the scheduling perspective through fine-grained operator fusion and the hardware perspective through design space exploration, using an extended version of the Stream modeling framework.

Our results demonstrate that the improved data locality stemming from our optimized fusion and scheduling strategy enables a speedup of up to 4.8× over unfused execution, while our adaptive memory-aware fusion approach reduces on-chip memory requirements by an order of magnitude without sacrificing performance. We further explore accelerator design trade-offs, showing that a fusion-aware hardware architecture can achieve 1.78× higher performance than the state-of-the-art MARCA accelerator, within the same area budget.
These results establish operator fusion as a key enabler for next-generation SSM accelerators.
\end{abstract}

%% file: 1-intro.tex
\begin{figure}[t]
    \centering
    \includegraphics[width=\linewidth]{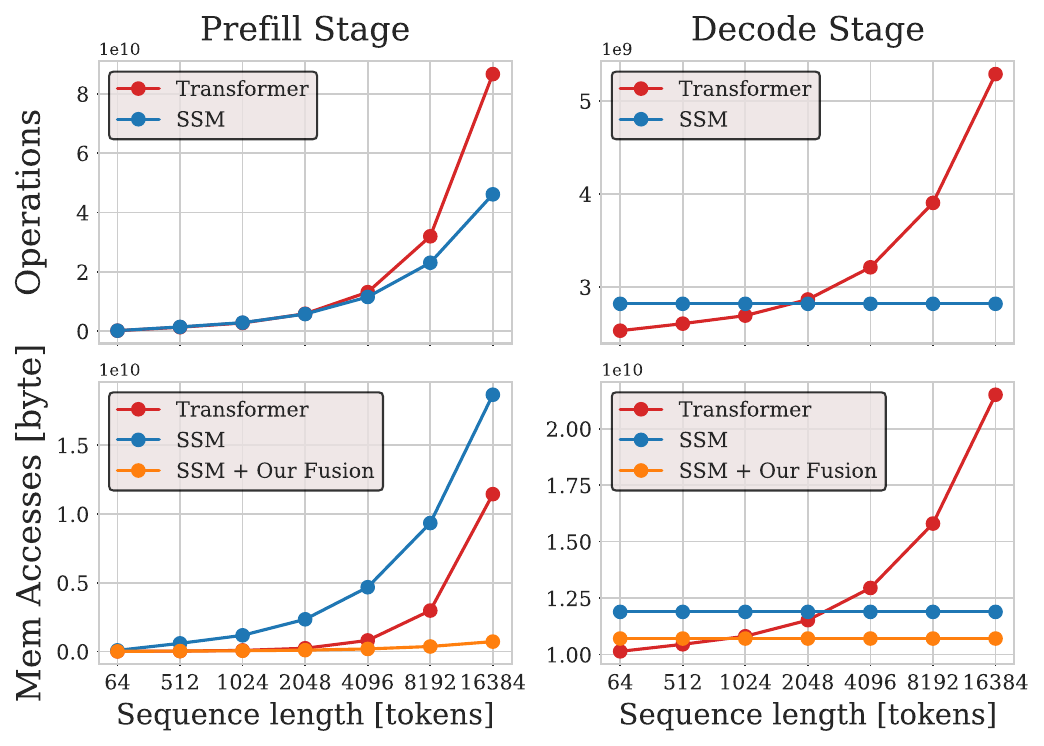}
    \caption{Comparison of transformer-based OPT-2.7B and SSM-based Mamba-2.8B during inference for the prefill stage (left) and decode stage (right). SSMs exhibit constant operations and memory usage during the decode stage, whereas the prefill stage presents a trade-off: lower operations but higher memory requirements compared to transformers.}
    \vspace{-4mm}
    \label{fig:ops_and_mem_accesses}
\end{figure}

\section{Introduction} \label{sec:intro}

Large Language Models (LLMs) with transformer-based architectures~\cite{attention-is-all-you-need} have revolutionized applications ranging from natural language processing to multimodal applications. Despite their immense success, deploying these models presents significant challenges due to 1) their enormous size, often exceeding a billion parameters, and 2) the quadratic computational and memory scaling of their self-attention mechanisms, which hinders the processing of long input sequences. Techniques such as caching, quantization, sparsity exploitation, and optimized dataflows~\cite{flat,fusemax} help mitigate these issues, but deployment challenges persist, particularly in resource-constrained environments and for long sequences.

State Space Models (SSMs)~\cite{gu2021combining,gu2021efficiently,mamba1} have emerged as a compelling alternative to transformers, offering linear computational complexity and constant memory scaling relative to sequence length.
Figure~\ref{fig:ops_and_mem_accesses} highlights this distinction by comparing computation and memory characteristics of transformers (OPT-2.7B~\cite{opt}) and SSMs (Mamba-2.8B~\cite{mamba1}) in two key processing stages. During the \textit{prefill stage}, when the model processes the initial input sequence before generating new tokens, SSMs require more frequent memory accesses compared to transformers, which poses challenges for efficient hardware acceleration. In contrast, during the \textit{decode stage}, where new tokens are generated sequentially, transformers rely on retrieving previously processed tokens from a key-value (KV) cache, while SSMs update a compact internal state instead. This gives SSMs a significant advantage for long sequences, reducing memory overhead and improving scalability.

\begin{figure}[tb]
    \centering
    \includegraphics[width=\columnwidth]{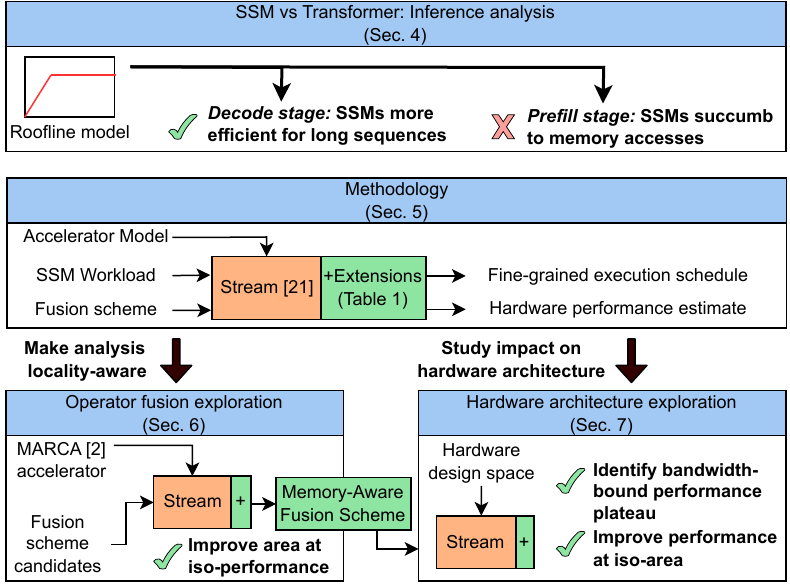}
    \vspace{-6mm}
    \caption{Overview of this paper.}
    \vspace{-4mm}
    \label{fig:framework}
\end{figure}

MARCA~\cite{li2024marca} represents the first effort towards dedicated hardware acceleration of SSMs, achieving up to $11\times$ speedup over a high-end GPU by employing highly-parallel reconfigurable processing elements. However, MARCA does not explore the impact of the number of parallel compute elements or the size of its embedded memory, presenting only a single design of an accelerator with a fixed area ratio of compute units to memory resources. The lack of a comprehensive analysis on the optimal hardware architecture configuration underlines the need for broader studies into SSM accelerator design, aiming to further enhance the performance of hardware accelerators for different use cases.

To address these challenges, this work comprehensively examines the computational characteristics of SSMs on dedicated hardware accelerators in order to provide broadly applicable insights. Our contributions are shown in Figure~\ref{fig:framework} and summarized as follows: 
\begin{itemize} 
    \item A high-level roofline analysis across transformers and SSMs, identifying the critical need for enhanced on-chip data reuse of SSMs during the prefill phase (Section~\ref{sec:roofline}). 
    \item Open-source extensions to the Stream modeling framework to enable precise performance estimation and automatic scheduling for SSM accelerators with optimized data locality (Section~\ref{sec:methodology}). 
    \item Development of an advanced operator fusion and scheduling strategy that can improve performance by as much as 4.8× (Section~\ref{sec:fusion}). 
    \item Exploration of the hardware accelerator design space showing that 1) identical performance constraints can be met while reducing the on-chip memory area by an order of magnitude, or 2) performance can be boosted by 1.78× under iso-area constraints compared to a state-of-the-art of SSM accelerator MARCA (Section~\ref{sec:hardware-dse}).
\end{itemize}

\cut{
By jointly addressing memory access patterns and architectural hardware inefficiencies in a coordinated way, this work lays the foundation for designing accelerators that optimize SSM performance across diverse use cases.}

%% file: 2-background.tex
\section{Background} \label{sec:background}

\subsection{LLM Inference Stages}
Large Language Models (LLMs) are deep neural networks—often consisting of hundreds of millions to hundreds of billions of parameters—that are trained on vast corpora of text data. 
During inference, LLMs generate new text predictions from a given prompt or input context.
Inference commonly involves two distinct computational stages: \textit{prefill} and \textit{decode}. During the prefill stage, the model processes the entire input sequence in one forward pass, computing sequence transformations that encode contextual relationships. In the subsequent decode stage, outputs are generated autoregressively, one token at a time. 
\cut{Different model architectures handle processing of the prefill and decode stages differently, which we discuss next for the transformer-based and SSM-based LLMs considered in this work.}

\subsection{The Dominance of Transformer-based LLMs}

Transformer-based LLMs~\cite{attention-is-all-you-need} have become fundamental to modern AI applications
\cut{, delivering state-of-the-art performance in natural language understanding, generation, and multimodal applications}. 
Their effectiveness largely stems from the self-attention mechanism that captures pairwise dependencies across tokens in an input sequence. However, this comes with a significant computational cost: during the \textit{prefill phase}
\cut{, where the model processes a full input sequence, }
self-attention operations scale as $\mathcal{O}(L^2)$ in time and in memory usage. 
\cut{In contrast, during the \textit{decode phase}, where tokens are generated one by one, the time complexity per token is reduced to $\mathcal{O}(L)$, }
This makes transformers particularly inefficient for long-sequence processing.

\subsubsection{Key Operations}

The computational workload of transformer models is dominated by two main types of operations, as illustrated on the left side of Figure~\ref{fig:model-architecture}:

\begin{figure}[tb]
    \centering
    \includegraphics[width=\linewidth]{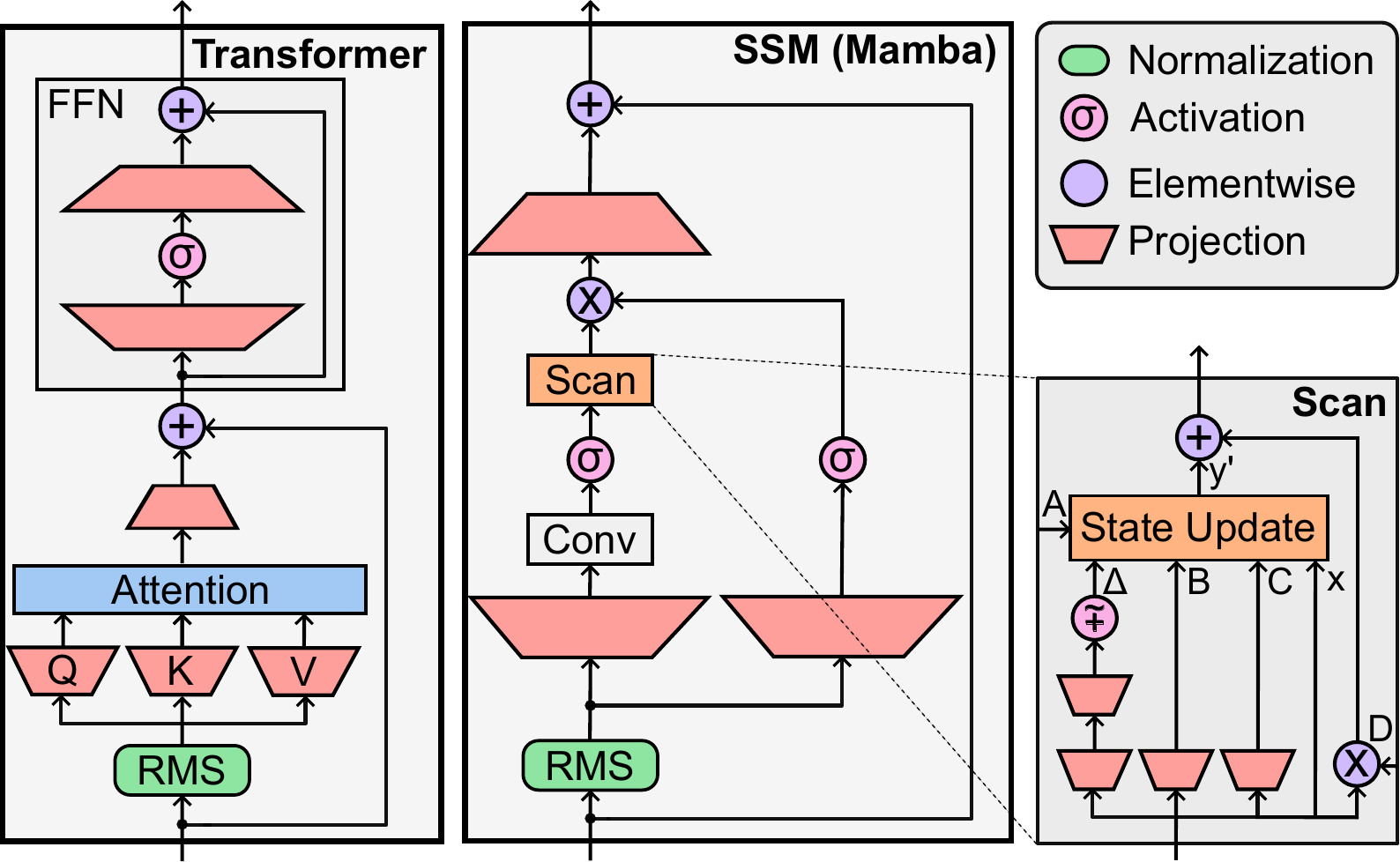}
    \vspace{-4mm}
    \caption{Architecture overview of transformers and SSMs. The state update block is detailed in Figure~\ref{fig:state-update}.}
    \label{fig:model-architecture}
\end{figure}

\textbf{Attention.}
The attention mechanism involves matrix multiplications to compute query-key similarity scores, followed by a softmax operation and matrix multiplication with value vectors to generate outputs. These operations collectively form the computational backbone of transformers, with the $\mathcal{O}(L^2)$ complexity arising from the computation of pairwise token interactions. Additionally, the softmax operation contains multiple non-linear operators and requires multiple passes over the input sequence, further increasing the implementation complexity.

\textbf{Projection.}
Projection operators are fully connected layers and are used to compute the $Q$, $K$, $V$ and attention output matrices, as well as in the feedforward network (FFN). These operators contain model weights that define the linear projection from the operator's input space to the output space. The weights of all projection layers collectively make up the transformer model's parameter count. 
\cut{The parameters are shared across tokens in the sequence, resulting in high data reuse for long sequences.}

\cut{
\textbf{Normalization.}
Normalization layers, such as the RMS layer normalization, are critical for stabilizing training and maintaining numerical stability during inference. These operators contribute little to the total operational count but have limited data reuse.
}

\cut{
\textbf{Elementwise Operations.}
Elementwise operations, including residual additions, are lightweight in terms of computation but have no data reuse.
}

\cut{
\textbf{Activation Functions.}
Nonlinear activations such as ReLU, GeLU or SiLU are applied within the feedforward network. The number of activation operations is low compared to attention and projection operators, but similar to the elementwise operations, they don't have any data reuse.
}

\subsubsection{Models}
Transformer-based LLMs have seen widespread development, with numerous models optimized for different use cases. One widely used model family in academic research is OPT~\cite{opt}, valued for its open-source availability and efficient scaling, making it a strong benchmark for studying transformer architectures. Other notable model families include GPT~\cite{gpt3}, known for its strong reasoning capabilities, and the open-source LLaMA~\cite{llama2}, which prioritizes deployment efficiency.


\subsubsection{Dedicated Accelerators}
In recent years, numerous hardware accelerators for transformer-based LLMs have been published \cite{Dai_Genc_Venkatesan_Khailany_2023,akaya,elsa,c-transformer,h3d-transformer,Keller_Venkatesan_Dai_Tell_Zimmer_Dally_Thomas,ita,quartet,hyctor}. These accelerators mainly target the prefill stage and employ custom techniques to exploit quantization or sparsity.

\subsection{State Space Models: A Promising Alternative}

State Space Models (SSMs) transform sequences by leveraging hidden state updates to retain past token information. Unlike recurrent neural networks (RNNs), which lack parallelized training, SSMs support efficient parallelization during training while maintaining sequential dependency modeling. At inference, they achieve $\mathcal{O}(L)$ (linear) computation and $\mathcal{O}(1)$ memory usage complexity.

\subsubsection{Key Operations}

The computational foundation of SSMs lies in the efficient use of the structured linear and elementwise transformations of a state-space representation. Figure~\ref{fig:model-architecture} (right) shows the model architecture of Mamba~\cite{mamba1}, a popular SSM. Key operations include:

\textbf{State Update.}
SSMs maintain a hidden state $h$ that serves as memory to process sequential data. In modern architectures such as Mamba, the state evolves through input-dependent transformation matrices, enabling dynamic updates. These updates involve multiple matrix-matrix multiplications, where the $D$-dimensional input is expanded by a factor $N$: the state dimension. The input $x$ is transformed with the input-dependent matrices after discretization (denoted as $\overline{A}$, $\overline{B}$, and $\overline{D}$) to produce an updated state and output $y$:
\vspace{-2mm}

\begin{equation}
\begin{aligned}
    h_{t} &= \overline{A} h_{t-1} + \overline{B} x_t \\
    y_t &= C h_t + \overline{D} x_t
\end{aligned}
\label{eq:ssm_discrete}
\end{equation}

\noindent One key characteristic of SSMs, and a popular argument to favor SSMs over more traditional RNNs, is the potential use of a parallel scan (\texttt{pscan}) algorithm~\cite{pscan}. The \texttt{pscan} algorithm updates the internal state in a parallel manner instead of a purely sequential one~\cite{Gu_Goel_Ré_2022}. This technique greatly speeds up training and inference time on GPUs by utilizing multiple hardware resources concurrently and eliminating sequential dependencies.

\textbf{Projection.}
The input is projected into a higher-dimensional $D$ space using input transformation matrices, while the output sequence is reconstructed through output transformation matrices.

\cut{
\textbf{Normalization.}
Similarly to transformers, SSMs include layer normalization, which is critical for stabilizing training and maintaining numerical stability during inference. These operators contribute little to the total operational count but have limited data reuse.
}

\textbf{Elementwise Operations.}
Elementwise operations, such as multiplications and residual additions, are computationally lightweight but lack data reuse, leading to inefficient memory access patterns. Despite their simplicity, their minimal data reuse can become a performance bottleneck, especially in bandwidth-limited hardware.

\cut{
\textbf{Activation Functions.}
Nonlinear activation functions, most commonly SiLU and SoftPlus, are applied to intermediate sequence representations. While the number of activation operations is relatively low compared to the state update and projection operators, they share the same limitation as elementwise operations: minimal data reuse.
}

\subsubsection{Models}
Early work on linear state spaces demonstrated the feasibility of representing sequences with structured time-invariant systems~\cite{gu2021combining}. Later, S4~\cite{gu2021efficiently} introduced parameter-efficient techniques to capture long-range dependencies, leveraging advanced initialization and diagonal state representations to rival transformers. Building on these ideas, Mamba~\cite{mamba1} further refined SSMs by introducing selective state spaces, optimizing different regions of the state for improved accuracy and computational efficiency. This has positioned Mamba as a strong competitor to transformers, particularly for long-sequence processing. Notably, it has been shown that Mamba achieves similar accuracy to the state-of-the-art transformer models when matched for parameter count, highlighting its potential as an efficient alternative~\cite{mamba1}.

A recent trend in LLM development is the emergence of hybrid transformer-SSM architectures that aim to leverage the strengths of both approaches. SSMs efficiently track long-term information in a more diffuse manner, akin to how humans recall the general atmosphere of past events, while transformers excel at retrieving precise details (like names or addresses). Notable examples include Zamba~\cite{zamba}, which stacks SSM and transformer layers sequentially, and Hymba~\cite{hymba}, which integrates both mechanisms within each layer by combining transformer and SSM heads in parallel. While this work primarily focuses on SSM acceleration, the presented techniques can also be applied to the SSM components of hybrid models.

\subsubsection{Dedicated Accelerators}

To the best of our knowledge, MARCA is the first and currently the only dedicated Mamba accelerator, designed for cloud environments. This architecture comprises $8192$ processing elements (PEs), \SI{24}{\mebi\byte} of on-chip SRAM, and \SI{256}{\giga\byte} of off-chip memory bandwidth, operating at \SI{1}{\giga\hertz}. This results in a design where \SI{80}{\percent} of the area is allocated to memory resources, leaving \SI{20}{\percent} for compute units.

Unlike traditional accelerators that rely on specialized functional units, MARCA employs flexible PEs capable of executing all required operations, including complex non-linear functions. This eliminates the need for separate functional units or vector arrays, minimizing area overhead. While certain operations require multiple cycles, they are still handled efficiently within the parallel PE arrays, maintaining computational versatility without additional architectural complexity.

\subsection{Operator Tiling and Fusion}  \label{sec:background-tiling}

Operator tiling optimizes on-chip data locality by restructuring loop-based computations to reduce data movement. For operators expressible as Einsums or nested loops, tiling partitions loop dimensions into outer and inner loops, ensuring that inner loops execute entirely within on-chip memory. This minimizes costly memory transfers, leading to improved performance. We define a \textit{tile} as the computation of all inner loops within a single outer loop iteration, while the \textit{number of tiles} corresponds to the total outer loop iterations required to process the full workload.

Operator fusion further enhances data locality by executing multiple operators consecutively while keeping intermediate results on-chip. A common example is fusing activation functions with preceding layers. Fusion is closely associated with tiling, where computations are first divided into tiles that fit within on-chip memory and then executed in a fused manner. For example, if two fused operators each have $n$ tiles, the execution proceeds in $n$ iterations, with each iteration processing the corresponding tiles from both operators in sequence. However, fusion requires knowledge of data dependencies between tiles, which is not always straightforward when intermediate tiles undergo complex transformations.

\begin{figure*}[t]
    \centering
    \includegraphics[width=\textwidth]{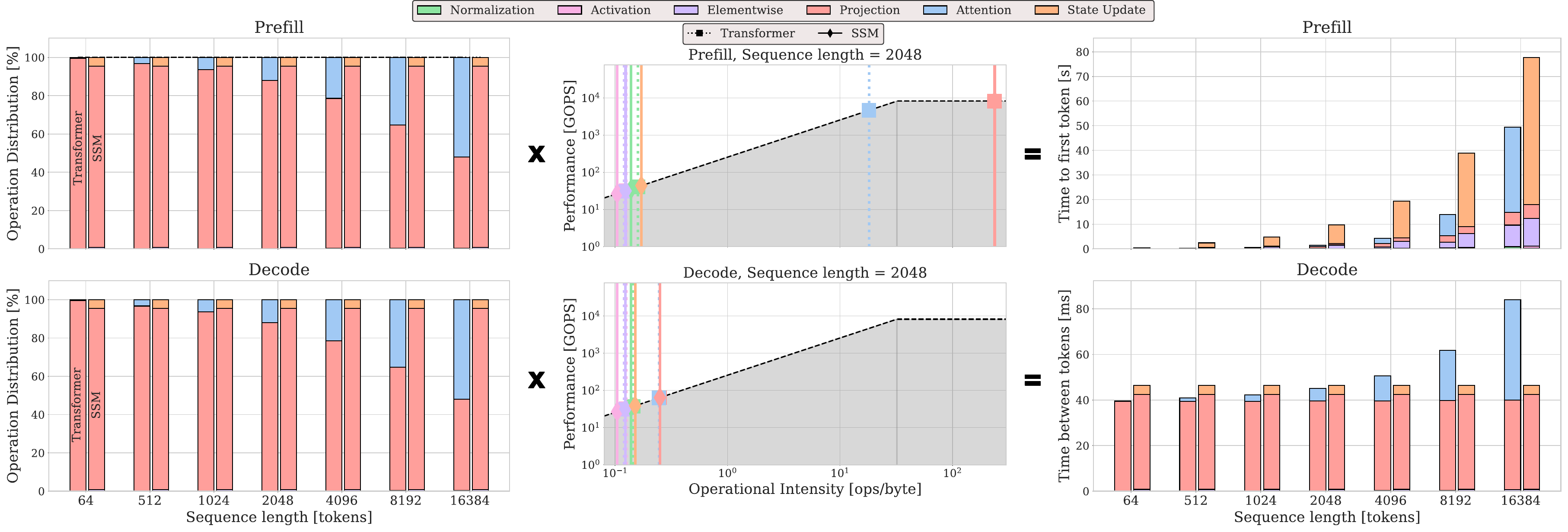}
    \vspace{-6mm}
    \caption{Inference latency (right) for the OPT-2.7B transformer (left bars, squares) and the Mamba-2.8B SSM (right bars, diamonds) across stages and sequence lengths. Operator distribution (left) and intensity (middle) shape the total latency. The roofline is shown for the $L\!=\!2048$ case. SSMs perform worse in the prefill stage due to memory-bound state-update operators.}
    \vspace{-3mm}
    \label{fig:roofline}
\end{figure*}

For transformer models, operator fusion is well-studied. Notably, FuseMax~\cite{fusemax} optimizes softmax computation by reducing the number of memory accesses and redundant passes over the sequence. FLAT~\cite{flat} improves on this work and fuses attention score computation, softmax normalization, and key-value weighting into a single operation, minimizing memory bandwidth requirements. However, for SSMs models, operator fusion studies are missing, aside from some GPU inference optimizations proposed by the author of Mamba~\cite{mamba1}.

\subsection{Stream Framework for Accelerator Modeling and Operator Fusion}

Stream~\cite{stream} is a modeling and scheduling framework that estimates and optimizes the performance of multi-core dataflow accelerators. Given an accelerator hardware architecture and a target neural network model for deployment, it models the on-chip dataflow, on-chip memory and core interconnection pattern with limited bandwidth to derive an optimized execution schedule and energy and latency predictions. Accurate performance estimation is obtained through a memory manager and a communication manager. The memory manager tracks the movement and storage of tensors, prioritizing on-chip data locality. The communication manager provides estimates of communication overheads for the required tensor transfers between cores. Stream supports a variable computation granularity—from coarse operator-by-operator processing to fine-grained operator fusion—and schedules all tiles with an adaptable scheduling order. This makes it a versatile tool for exploring both the fusion and accelerator design space. However, while Stream supports a wide range of different workloads through an ONNX~\cite{onnx} front-end, there are key aspects missing for modeling LLMs.

%% file: 3-motivation.tex
\section{Motivation, Challenges \& Goals} \label{sec:motivation}

\subsection{Motivation}
Efficient LLM inference is critical for many applications, particularly in real-time or autonomous systems that demand low-latency execution. Custom hardware accelerators have played a key role in optimizing efficiency for domain-specific workloads, particularly in deep learning. While numerous accelerators have been proposed for transformer-based LLMs, far fewer exist for SSM-based models, despite their promising efficiency in long-sequence processing.

However, developing specialized accelerators is a long and costly process. Hardware design cycles can take months, making it essential to gain early insights into viable accelerator architectures.

\subsection{Challenges}
We identify three key challenges for efficient LLM hardware acceleration:

\inlineheading{Computational Complexity}
LLMs require extensive computation and data movement. However, the workload characteristics vary drastically depending on the use case, such as inference stages (prefill, decode) and sequence length. It is unclear which LLM architecture performs best when different scenarios are considered, as demonstrated in Figure~\ref{fig:ops_and_mem_accesses}. A deep understanding of operator types and their behavior on accelerators is necessary to optimize inference performance.

\inlineheading{Operator Fusion}
Reducing data movement through operator fusion has been widely studied for transformer-based models~\cite{fusemax, flat}. However, the potential for operator fusion in SSMs remains largely unexplored. Understanding which SSM operators can be fused, what tile sizes are optimal, and how this impacts hardware performance is essential for improving efficiency.

\inlineheading{Rapid Hardware DSE}  
LLM architectures evolve rapidly, making it crucial to assess their impact on accelerator performance early. Key hardware components—including compute resources, dataflow, memory capacity, and off-chip bandwidth—all influence cost and performance. MARCA~\cite{li2024marca}, a state-of-the-art SSM accelerator, provides insights into one specific design point but does not explore the broader design space. Traditionally, evaluating different architectural configurations requires multiple low-level hardware implementation iterations, each taking weeks, while cycle-accurate simulations of LLM models with high parameter counts further slow down exploration by requiring days per evaluation. Faster evaluation methods are needed to efficiently analyze the design space and guide architectural decisions beyond a single accelerator instance.

\subsection{Goals}
The goals of this paper link to the three challenges:

\inlineheading{Computational Complexity} Analyze the operator distribution in transformer- and SSM-based LLMs and perform a high-level performance analysis across different use cases.

\inlineheading{Operator Fusion} Explore new fusion and scheduling strategies to optimize on-chip data reuse and efficiently accelerate SSM models.

\inlineheading{Rapid Hardware DSE} Evaluate how hardware architecture decisions impact SSM inference performance, incorporating optimized fusion strategies to come to improved accelerator architectures.

\section{Transformer \lowercase{vs.} SSM Analysis} \label{sec:roofline}

Figure~\ref{fig:ops_and_mem_accesses} summarizes the total operations and memory accesses required for inference across different stages (prefill and decode) of transformer- and SSM-based LLMs, providing proxy metrics for performance. Throughout this paper, we use OPT-2.7B~\cite{zhang2022opt} as the representative transformer model and Mamba-2.8B~\cite{mamba1} as the SSM model. These models are well-matched in terms of parameter count and accuracy~\cite{mamba1}, making them a fair basis for comparison. However, the total number of operations alone does not fully capture latency bottlenecks, as different operator types contribute unequally to execution time.

\subsection{Operation Distribution}

To better understand the unequal contribution of operator types to the total latency, Figure~\ref{fig:roofline} (left) shows the distribution of the number of operations across different operator types for transformers (left bars) and SSMs (right bars). Notably, these distributions remain nearly identical between the prefill and decode stages. For transformers, this is due to KV caching, which reduces both projection and attention computations proportionally. For SSMs, the distribution remains unchanged across stages and sequence lengths due to their recurrent formulation, where structured state-space updates and projections are applied uniformly at each step.

Among all operator types, projections account for the largest share of operations, except in transformers at large sequence lengths, where attention begins to dominate due to the quadratic scaling of query-key interactions.

\subsection{Operational Intensity}
The latency cost of operator types is heavily influenced by their \textit{operational intensity} ($OI$), defined as the ratio of operations to memory accesses (in operations per byte). A higher $OI$ indicates better data locality (cfr. Section~\ref{sec:background-tiling}), as more computations are performed per byte transferred from memory. The x-axis of Figure~\ref{fig:roofline} (middle) illustrates the $OI$ of different operator types in transformers (dashed lines, squares) and SSMs (full lines, diamonds).

In the prefill stage, projection operators of both transformers and SSMs exhibit the highest $OI$ due to their large matrix-matrix multiplications, which enable significant data reuse. Transformer attention also exhibits high $OI$, as the query-key multiplication operates on the entire sequence at once, leading to substantial reuse of key and query activations. This results in an $OI$ of 18.1 ops/byte. In contrast, SSM state updates apply structured recursions on a single token at a time, preventing any substantial reuse across the sequence. As a result, their $OI$ is significantly lower at just 0.17 ops/byte.

In the decode stage, $OI$ drops significantly across all operators. Unlike in the prefill phase, where computation on the entire sequence enables data reuse, the decode stage operates autoregressively, processing one token at a time. This severely limits data locality, forcing frequent memory accesses for each new token. As a result, even projection operations, which were previously compute-bound, shift toward a memory-bound regime. This fundamental characteristic impacts both transformers and SSMs.

\subsection{Roofline Model Performance}

The \textit{hardware roofline model} provides an intuitive framework to analyze the performance of different operators (characterized by their $OI$), relative to the hardware specifications in terms of off-chip bandwidth and peak computational performance~\cite{williams2009roofline}, and identifies memory-bound and compute-bound regions. 
\cut{It characterizes the achievable performance \( P \) (in GOPS) as a function of $OI$, revealing whether an operation is compute-bound or memory-bound. The model is defined as:}

\cut{
\[
P = \min\left(P_\text{peak}, BW \cdot OI\right),
\]
}

\cut{
where \( P_\text{peak} \) is the theoretical peak performance of the hardware (in GOPS), and \( BW \) represents the peak memory bandwidth (in GB/s). When $OI < \frac{P_\text{peak}}{B}$, performance is limited by memory bandwidth (memory-bound regime). Conversely, when \( I > \frac{P_\text{peak}}{B} \), performance saturates at \( P_\text{peak} \) (compute-bound regime).
}

Figure~\ref{fig:roofline} (middle) presents the roofline plot for the MARCA accelerator with its $P_\text{peak}\!=8192 \ GOPS$ and $BW\!=256 \ GB/s$. We superimpose its impact on transformer and SSM-based models for a sequence length of \( 2048 \) tokens. The performance of each operator type is computed based on its $OI$ and the hardware constraints. As expected, projection operators fall in the compute-bound region due to their high reuse. Meanwhile, all other operators, including attention and SSM state updates, remain memory-bound. Despite both being memory-bound, attention achieves a performance of $4633 \ GOPS$, compared to the SSM state-update $44 \ GOPS$, a performance reduction of more than $100\times$.

\subsection{Latency Implications}
To approximate inference latency, we combine operation counts with the operational intensity ($OI$) of each operator. This estimate assumes a conventional layer-by-layer execution model, without leveraging operator tiling or fusion optimizations. Figure~\ref{fig:roofline} (right) illustrates the absolute latency for both models. For the prefill stage, latency corresponds to the time required to generate the first token (measured in seconds, shown at the top). In the decode stage, it represents the time between consecutive token generations (measured in milliseconds, shown at the bottom).

\subsubsection{Prefill Stage}
Although projection operators dominate the operation count, their high $OI$ ensures they have minimal impact on latency. The latency of the transformer scales roughly quadratically with sequence length due to the quadratic growth of attention operations, while the SSM’s latency increases linearly. However, despite its favorable scaling, the SSM exhibits consistently higher latency than the transformer across all sequence lengths. This is primarily due to the memory-bound SSM state update operations that suffer from an $OI$ approximately $100\times$ lower than the attention of the transformer. 

\inlineheading{Takeaway 1} The SSM's latency is dominated by state update operations with $100\times$ lower $OI$ compared to transformer's attention operators.
By increasing the $OI$ of these operators through data reuse optimizations, we can further reduce the total latency. In the next section, we explore how operator fusion can be leveraged to achieve this improvement.

\subsubsection{Decode Stage}
In the decode stage, the SSM maintains constant latency across sequence lengths due to the fixed size of its internal state. For short sequences, it is slower than the transformer because of a large state expansion factor. However, its latency is dominated by projection operators, which are relatively slow.
In contrast, the transformer's latency increases with sequence length because attention operations scale linearly in this stage.\footnote{Attention grows linearly, not quadratically, with sequence length in the decode stage since only a single token is generated at a time.} Consequently, for long sequences, the SSM is significantly faster, as intended by design.

\inlineheading{Takeaway 2} In the decode stage, SSM latency is fundamentally constrained by the dominance of projection operators and memory bottleneck. Further latency reductions would require either increased bandwidth, reduced bitwidth of the projection weights, or alternative accelerator technologies, such as digital in-memory computing.

%% file: 3.1_methodology.tex
\section{Methodology} \label{sec:methodology}

This section presents the main methodology used throughout the rest of this paper, as illustrated in Figure~\ref{fig:framework}. At its core, we employ Stream~\cite{stream}, a versatile performance estimation framework capable of tracking memory contents over time.

While the roofline analysis offers high-level insights into computational bottlenecks and memory-boundedness, it does not capture on-chip data locality across operators or account for resource utilization. By leveraging Stream, we overcome these limitations, enabling a more detailed investigation of on-chip memory utilization, tiling, and fusion strategies. This analysis is particularly crucial during the prefill phase of SSM execution, where efficient memory access patterns play a key role in overall performance.

Section \ref{sec:stream-extensions} proposes the open-source extensions to Stream that enable flexible tiling of multiple SSM operators, accurate identification of inter-tile dependencies, and efficient scheduling of fused tiles. Next, the precise setup and major inputs to Stream are specified: Section \ref{sec:workload-model} details how the SSMs are translated to ONNX format and Section \ref{sec:hardware-model} defines the high-level hardware accelerator model used throughout the further explorations and optimizations of this paper.

\subsection{Extending Stream for Dependency-Oriented Analysis} \label{sec:stream-extensions}

\begin{figure}[tb]
    \centering
    \includegraphics[width=\linewidth]{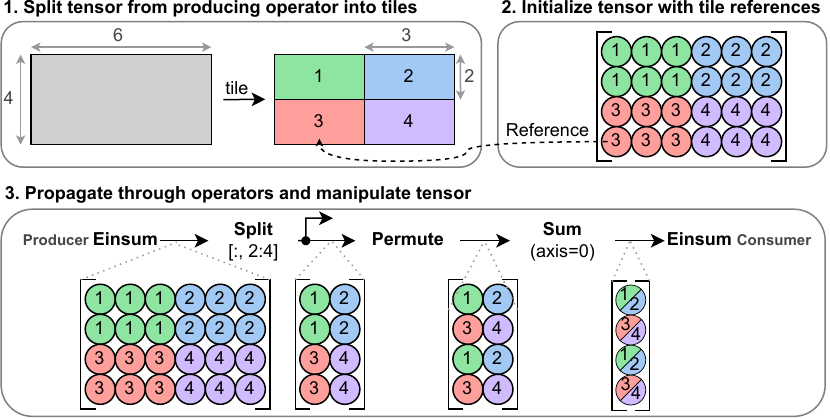}
    \vspace{-6mm}
    \caption{Dependency tracking using emulation of complex tensor manipulations between a \textit{producing} and \textit{consuming} operator. At the end, the exact dependencies for every individual element in the consuming operator's input are known.}
    \vspace{-4mm}
    \label{fig:nodetensor}
\end{figure}

To accurately represent SSM workloads and track dependencies between fused tiles, we introduce several modifications to the Stream framework. The extensions, summarized in Table~\ref{tab:stream-extensions} and detailed next, will be made available as open-source contributions.

\begin{table}[h!]
    \centering
    \caption{Our extensions to Stream framework}
    \vspace{-3mm}
    \setlength{\tabcolsep}{3pt} 
    \begin{tabular}{m{0.22\linewidth} >{\centering\arraybackslash}m{0.22\linewidth} >{\centering\arraybackslash}m{0.42\linewidth}}
        \toprule
        & \textbf{Stream \cite{stream}} & \textbf{Ours} \\ 
        \midrule 
        \textbf{Operator \newline Support} & GeMM, Conv, \newline Elementwise          & +Einsum,  Softmax, \newline External product, Reduction  \\
        \hline
        \textbf{SSM State \newline Update} & \ding{55} & \ding{51} \\
        \hline
        \textbf{Dependency \newline Tracking} & R-tree &  Tensor-based (Fig.~\ref{fig:nodetensor}) \\
        \hline
        \textbf{Multi-cycle \newline Operators} & \ding{55} & \ding{51} \\

        \bottomrule
    \end{tabular}
\label{tab:stream-extensions}
\end{table}

\subsubsection{Operator Support}
While Stream accepts ONNX models as workload inputs, standard ONNX operators are insufficient to represent the state update mechanism of SSMs due to its recurrent nature. To address this limitation, we extend the ONNX operator set with a custom \texttt{SSM} operator and modify Stream’s ONNX parser to correctly interpret this operator. The parser decomposes the \texttt{SSM} operator into atomic operations, while also ensuring proper handling of dependencies between the atomic operations, and between prior operators and the state update block. The implementation of the state update is later detailed in Section~\ref{sec:workload-model}. 

Other SSM that were not yet supported by Stream's ONNX parser, such as Einsum operations, external products, and reduction operations (e.g., the sum over a given axis), are also added.

\subsubsection{Dependency Tracking}
We propose a novel dependency tracking mechanism that efficiently and accurately infers all inter-tile dependencies, regardless of the tiling and fusion strategy. The existing Rtree-based dependency tracking in Stream is limited to cases where producer and consumer tiles have matching dimensions. However, SSMs frequently employ operators such as \texttt{Split}, \texttt{Slice}, and \texttt{Transpose}, that alter the tensor dimensions between the producer and consumer tiles. To accommodate these transformations, a more general dependency tracking mechanism is required.

To address this challenge, we introduce a tensor-based approach that tracks dependencies at element granularity, as summarized in Figure \ref{fig:nodetensor}. In this method, the output of each (untiled) operator is represented as a tensor, where each element references the tile that produced it. These dependency-tracking tensors enable arbitrary tiling strategies as specified by the user. They are then propagated through the complex operators, accurately reflecting operations that modify the shape, order, or structure of elements.
\cut{
Due to the large intermediate tensors involved in SSM workloads, maintaining this level of dependency tracking can be computationally expensive. To mitigate this overhead, we optimize the implementation by leveraging NumPy intrinsics for efficient tensor manipulation.}
Additionally, we introduce heuristics to identify tensor dimensions unaffected by specific transformations and exclude them from dependency tracking, thereby reducing unnecessary computations. 


\begin{figure}[t!]
    \centering
    \includegraphics[width=0.9\linewidth]{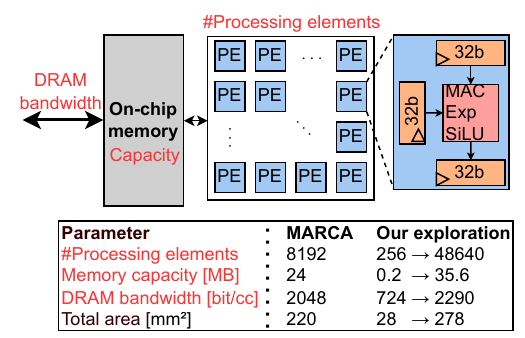}
    \vspace{-3mm}
    \caption{Diagram of the parameterized accelerator model, with parameter values for MARCA and our hardware design space exploration (Sec.~\ref{sec:hardware-dse}).}
    \label{fig:marca-architecture}
    \vspace{-3mm}
\end{figure}

\subsubsection{Multi-cycle Operators}
Beyond the standard MAC operations, SSMs incorporate operators that may require multiple cycles per operation when executed on hardware accelerators, such as exponentiation and activation functions.  This characteristic affects achievable performance, potentially increasing latency in compute-bound scenarios while having minimal impact on memory-bound computations. To accurately capture this effect, we extend Stream to allow users to specify cycle-per-operation ($CPO$) characteristics for individual operators. 
\cut{
By incorporating these annotations, we improve performance estimation and provide deeper insights into the relationship between operator latency and overall workload efficiency.}

\subsection{Workload model} \label{sec:workload-model}
Stream accepts an ONNX~\cite{onnx} graph as input to represent workloads. The full SSM shown in Figure~\ref{fig:model-architecture} is first implemented in PyTorch and then exported to ONNX. By leveraging our extensions to the ONNX parser, Stream can correctly interpret all SSM operations.

As mentioned in Section~\ref{sec:background}, the state update block is typically implemented using a \texttt{pscan} algorithm. However, in this work, we opt for a purely sequential implementation for three reasons:
1) Dataflow architectures with 2D PE arrays do not benefit as much from \texttt{pscan} as the streaming multiprocessors found in GPUs.
2) \texttt{pscan} necessitates batched execution for large sequence lengths due to its substantial memory requirements.
3) \texttt{pscan} has a higher time complexity of $\mathcal{O}( L \log_2 L)$~\cite{nvidia-pscan}, whereas the sequential implementation has a linear complexity of $\mathcal{O}(L)$. 

The decomposition of the sequential state update block is illustrated in Figure~\ref{fig:state-update}. In this implementation, the  $\Delta$, $B$ and $x$ tensors are first processed for all timesteps simultaneously, yielding the tensors $Exp(A)$ and $\Delta B x$. The state $h$ is then updated sequentially for each timestep, starting from the initial state $h_0$. At each timestep, only a single slice of the $Exp(A)$ and $\Delta B x$ tensors is used, extracted by indexing the $L$ dimension with the timestep index. At any given time, only one intermediate version of the state is stored, yet all $y'$ tensors (one per timestep) are retained for later use.


\subsection{Hardware Model} \label{sec:hardware-model}

Stream requires a high-level description of the target hardware accelerator as input. In this abstraction, an accelerator consists of multiple cores connected via a user-defined interconnect topology, where each link is characterized by its bandwidth and energy cost. A core may represent either a computational unit, defined by its on-chip memory hierarchy and processing element (PE) array dataflow, or a memory unit such as off-chip working memory. This flexible representation enables the modeling of heterogeneous architectures with diverse core configurations.

For our analysis, we explore SSM-specific accelerators in Stream starting from MARCA~\cite{li2024marca}. Figure~\ref{fig:marca-architecture} illustrates the accelerator template alongside the parameter values used to model MARCA, as well as the parameter ranges explored in Section~\ref{sec:hardware-dse}. In correspondence with MARCA's PE design, we set the $CPO$ of exponentiation, SiLU and sigmoid operations to 4. Given that our primary focus is the impact of on-chip data locality through operator fusion on overall latency, we assume a sufficiently high on-chip memory bandwidth to prevent computational stalls in the PE array. This assumption is reasonable for accelerators, as their PE arrays and memory hierarchies are typically co-optimized. To further expedite simulations, MARCA's 32 tensor cores are aggregated into a single core with $8192$ PEs, while still accounting for potential spatial under-utilization. 
These specifications align with contemporary accelerator designs, providing a representative framework for deriving broadly applicable insights.

For the workloads and architectures studied in this paper, the simulation time for a single design point is in the order of minutes.

%% file: 4-fusion.tex
\section{Architecture- \& Fusion-aware Analysis} \label{sec:fusion}

\begin{figure}[tb]
    \centering
    \includegraphics[width=0.9\linewidth]{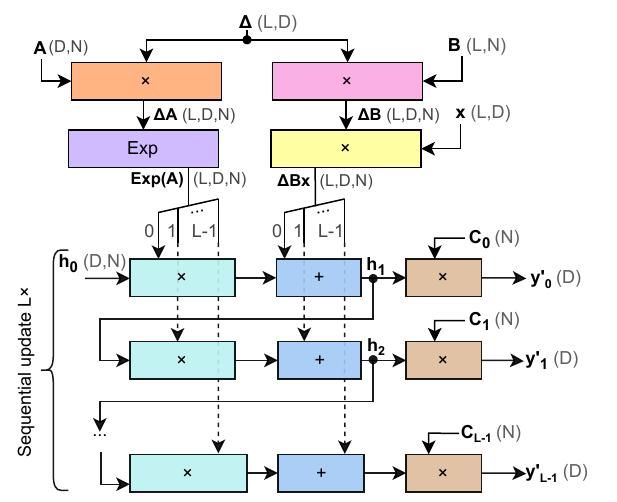}
    \vspace{-2mm}
    \caption{Diagram of the sequential state update block for $L$ timesteps. Tensor names are in bold.}
    \label{fig:state-update}
    \vspace{-3mm}
\end{figure}

\begin{figure*}[t]
    \centering
    \includegraphics[width=0.9\linewidth]{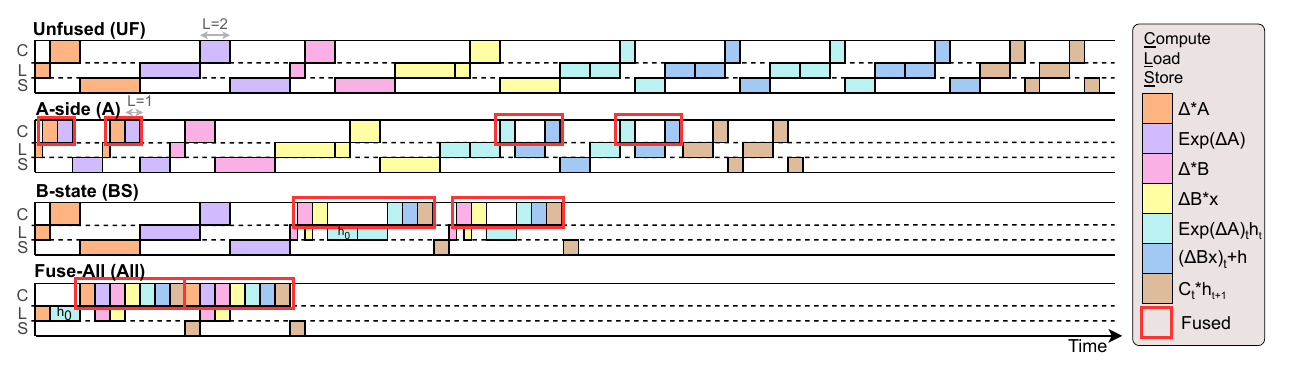}
    \vspace{-5mm}
    \caption{Execution schedule of different fusion schemes. Not drawn to scale.}
    \label{fig:schedule}
    \vspace{-2mm}
\end{figure*}

In this section, we leverage Stream to explore various operator fusion strategies aimed at improving on-chip data locality. After exploring the fusion depth in Section \ref{sec:fusion-depth}, we analyze the hardware implications of the most effective fusion scheme in Section \ref{sec:memory-size}, followed by further optimizations in Section \ref{sec:split-D}.

\subsection{Exploring Fusion Depth} \label{sec:fusion-depth}
To incorporate the inter-operator data reuse opportunities in the SSM analysis, we devise various fusion possibilities and evaluate their effect on latency. As a first step, we explore the effect of the fusion depth, defined as the number of consecutive operator tiles executed without reloading intermediate activations from off-chip memory. 

\begin{table}[b]
    \caption{Fusion schemes under study. Tensors are named in correspondence with Figure~\ref{fig:state-update}. $n$ is computed at compile time, depending on the on-chip memory capacity.}
    \vspace{-2.5mm}
    \label{tab:fusion-schemes}
    \begin{tabular}{llll}
    \hline
    \textbf{\begin{tabular}[c]{@{}l@{}}Fusion \\ scheme\end{tabular}} &
      \textbf{Abbrv.} &
      \textbf{Locality of tensors} &
      \textbf{\begin{tabular}[c]{@{}l@{}}\#tiles per \\ fused layer\end{tabular}}  \\ \hline
    \texttt{Unfused}           & \texttt{UF}   & None    & None       \\
    \texttt{A-side}            & \texttt{A}    & $\Delta A$ & $L$      \\
    \texttt{B-side}            & \texttt{B}    & $\Delta B$  & $L$       \\
    \texttt{AB-sides}          & \texttt{A-B}  & $\Delta A$, $\Delta B$ & $L$     \\
    \texttt{A-state}           & \texttt{AS}   & $\Delta A$, $Exp(\Delta A)$, $h$ ($\times 2$)  & $L$     \\
    \texttt{B-state}           & \texttt{BS}   & $\Delta B$, $\Delta Bx$, $h$ ($\times 2$)               & $L$   \\
    \texttt{A-state,B}         & \texttt{AS-B} &  $\Delta A$, $\Delta B$, $h$ ($\times 2$), $\Delta B$   & $L$   \\
    \texttt{B-state,A}         & \texttt{BS-A} & $\Delta B$, $\Delta Bx$, $h$ ($\times 2$), $\Delta A$  & $L$    \\
    \texttt{Fuse-All}          & \texttt{All}  & All of the above & $L$   \\
    \texttt{Mem-Aware}     & \texttt{MA-All}   & All of the above & $n L$  \\ \hline
    \end{tabular}
\end{table}

We limit the fusion evaluation to operators within the SSM state update block depicted in Figure~\ref{fig:state-update}, as they are expected to benefit the most from operator fusion due to their low $OI$, as discussed in Section~\ref{sec:roofline}. 
Each fusion scheme is defined by a specific set of fused operators and corresponding on-chip tensor retention strategies. 

We tile output tensors along the token dimension $L$, effectively splitting $(L, D, N)$ tensors into $L$ tiles of shape $(1, D, N)$.
Tiling along the $L$ dimension ensures compatibility with the sequential state update and identical performance independent of sequence length. In contrast, tiling along the $D$ or $N$ dimension alone would yield tensor sizes that still depend on $L$, potentially exceeding on-chip memory limits for long sequences and negating fusion benefits.



Table~\ref{tab:fusion-schemes} summarizes all explored fusion schemes, where the number of local on-chip tensors corresponds to the fusion depth. Each listed tensor is tiled into $L$ parts, with tiles scheduled consecutively to ensure immediate consumption by subsequent operators, preventing off-chip memory transfers. Figure~\ref{fig:schedule} illustrates this process for a simplified case where $L\!=\!2$.

\begin{figure}[t]
    \centering
    \includegraphics[width=\linewidth]{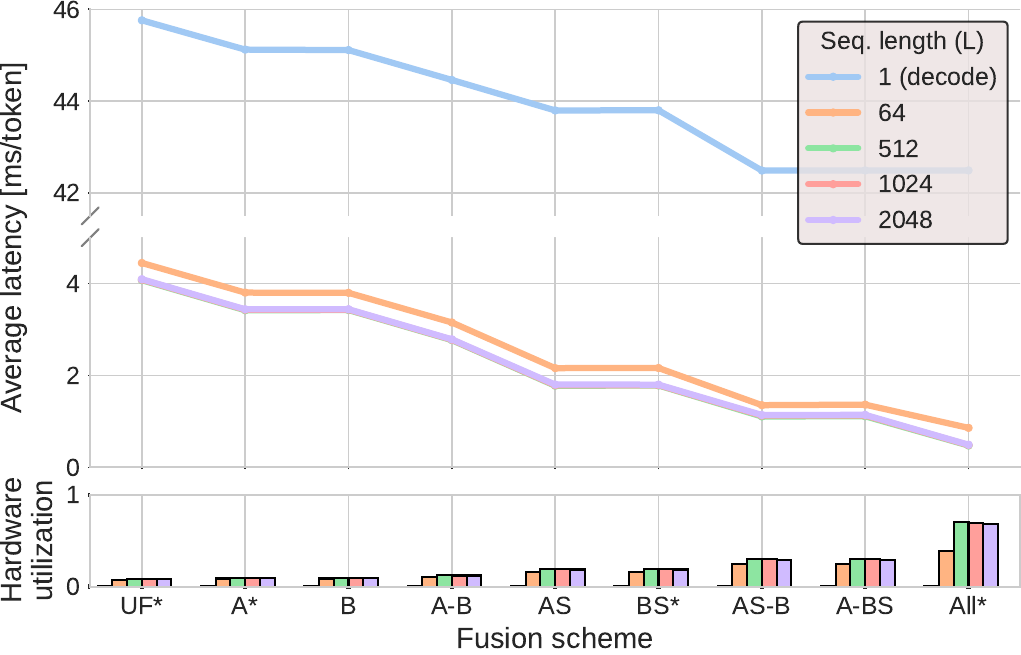}
    \vspace{-7mm}
    \caption{Evaluation of different fusion schemes using Stream. Asterisk($^*$) indicates fusion schemes are visualized in Figure~\ref{fig:schedule}.}
    \vspace{-8mm}
    \label{fig:fusion-depth}
\end{figure}

Figure~\ref{fig:fusion-depth} compares the estimated latency (normalized by the sequence length) and the overall hardware utilization for the entire SSM model across all fusion schemes and for different sequence lengths. Each increase in fusion depth consistently reduces latency due to reduced off-chip data transfers, with similar latency reductions observed for schemes of equal depth. The lowest latency is achieved when all state update operators are fused, eliminating all off-chip memory accesses. This scheme, denoted as \texttt{Fuse-All}, achieves an average speedup of 4.8× over unfused execution for long sequences. Under \texttt{Fuse-All}, the state update module reaches an average hardware utilization of $98.3\%$ for all sequence lengths, effectively shifting from a memory-bound to a compute-bound regime.

Note that the average latency per token of inferences with $L\!=\!1$ (decode phase) is significantly higher due to memory-bound projection operators. For larger sequences ($L \ge 512$ in Figure~\ref{fig:fusion-depth}), the average latency per token becomes independent of $L$. For these values, the projections are compute-bound, and the total latency scales linearly in $L$, keeping the average latency per token constant. This demonstrates that our analysis remains valid for even longer sequences.

\inlineheading{Takeaway 3} By tiling and fusing all operators within the state update block in the $L$ dimension, its computations shift from a memory-bound to a compute-bound regime, greatly improving performance.

\subsection{Sensitivity to On-chip Memory Capacity} \label{sec:memory-size}

The \texttt{Fuse-All} scheme eliminates all off-chip memory transfers for the MARCA-based accelerator model with \SI{24}{\mebi\byte} of on-chip memory. Next, we investigate the impact of reduced on-chip memory on performance.

\begin{figure}[t!]
    \centering
    \includegraphics[width=0.75\linewidth]{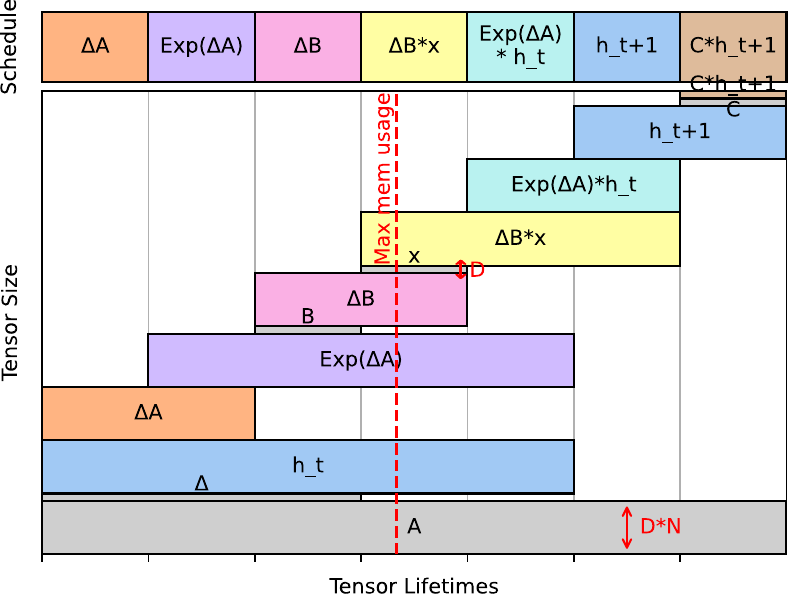}
    \vspace{-2mm}
    \caption{Schedule of operators (top) and lifetimes of tensors (bottom) of one state update iteration. $A$ and $h$ remain in memory throughout the entire fused execution as they are reused across iterations.}
    \label{fig:tensor-lifetime}
    \vspace{-4mm}
\end{figure}

The peak memory usage during fused execution can be derived analytically in terms of SSM dimension sizes. Figure~\ref{fig:tensor-lifetime} illustrates the tensor lifetimes for a single state update timestep under the \texttt{Fuse-All} scheme, with peak memory demand occurring during the computation of $Exp(\Delta A)_t \cdot h_t$. At this point, storage is required for five $(D, N)$-size tensors and one $(D,)$-size tensor. Therefore, \texttt{Fuse-All} sets the following requirement on the on-chip memory capacity, assuming 32-bit activations:
\vspace{-4mm}

\begin{equation} \label{eq:mem-size}
   |Memory| > (5DN+D) \times 32 \ bit
\end{equation}

To test this hypothesis, we gradually reduce the on-chip memory capacity of our accelerator model and estimate the inference latency for the \texttt{Fuse-All} scheme, shown with full lines in Figure \ref{fig:memory-sweep-merged}. The latency exhibits a staircase pattern, where each step corresponds to an additional off-chip transfer when an intermediate tensor no longer fits on-chip. The first major latency increase aligns with the threshold predicted by Equation~(\ref{eq:mem-size}).


\inlineheading{Takeaway 4} Under fusion scheme \texttt{Fuse-All}, the on-chip memory capacity of contemporary SSM accelerator architectures can be significantly reduced without sacrificing performance, up to the threshold of Equation~(\ref{eq:mem-size}).

\subsection{Memory-Aware Fusion} \label{sec:split-D}

To enable memory-efficient inference on smaller accelerators or of SSMs with larger dimensions, we introduce a finer-grained adaptive fusion scheme: instead of solely tiling along the $L$ dimension, activation tensors are further partitioned along the $D$ dimension.

We choose $D$, as partitioning the $N$ dimension into many small tiles would lead to spatial under-utilization of the PE array. For instance, in Mamba1-2.8B, $N\!=\!64$ while $D\!=\!5120$.

Starting from Equation~(\ref{eq:mem-size}), we determine the minimum number of splits $n$ required in the $D$ dimension to prevent off-chip transfers in the \texttt{Fuse-All} scheme, as follows:
\vspace{-2mm}

\begin{equation} \label{eq:D-split}
   n = \left\lceil{ \frac{(5DN+D) \times 32 \ bit}{|Memory|} }\right\rceil 
\end{equation}

\begin{figure}[t!]
    \centering
    \includegraphics[width=\linewidth]{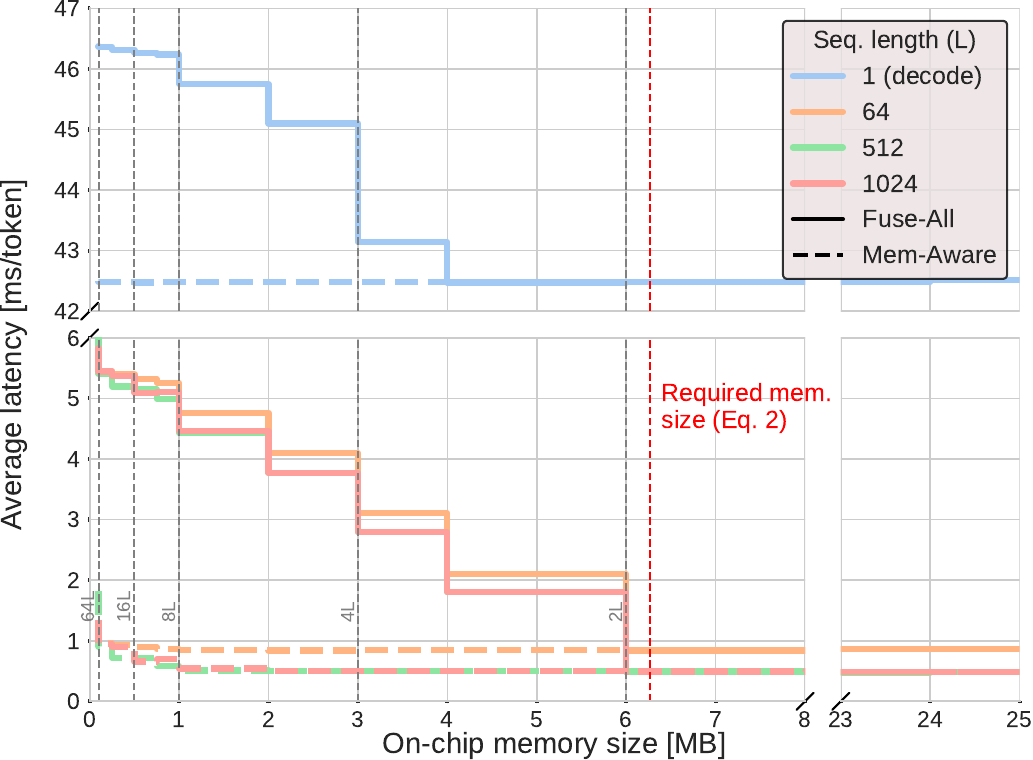}
    \vspace{-5mm}
    \caption{Sensitivity of latency to on-chip memory capacity under fusion scheme \texttt{Fuse-All} (full lines) and \texttt{Mem-Aware} (dashed lines). For \texttt{Fuse-All}, all fused tensor are split into $L$ tiles. For \texttt{Mem-Aware}, the number of tiles per tensor is indicated next to the gray dashed lines.}
    \vspace{-6mm}
    \label{fig:memory-sweep-merged}
\end{figure}

The fusion scheme that results from tiling the fused operators in both the $L$ dimension ($L$ times) and $D$ dimension ($n$ times) is denoted as \texttt{Mem-Aware}. Figure~\ref{fig:memory-sweep-merged} illustrates the performance impact of \texttt{Mem-Aware} fusion in function of memory capacity with dashed lines. The results demonstrate that inference latency remains stable even with a 24× smaller on-chip memory. These findings suggest that large on-chip memory capacities, such as the 80\% silicon area allocation of MARCA's \SI{222}{\milli\meter\squared} die, may be unnecessary for achieving optimal performance.

\inlineheading{Takeaway 5} Under the fine-grained \texttt{Mem-Aware} fusion scheme, the on-chip memory capacity of contemporary SSM accelerator architectures can be reduced by an order of magnitude without sacrificing performance.

%% file: 5-dse.tex

\section{Architecture Design Space Exploration} \label{sec:hardware-dse}
  
Section~\ref{sec:fusion} concludes that the accelerator’s on-chip memory capacity, and consequently its area, can be substantially reduced under the same latency constraints. This finding prompts our final study: given a specific inference scenario and area budget, what constitutes an optimal SSM accelerator design, i.e. how should computational area be balanced against memory area? To answer this, we explore the accelerator design space, leveraging our enhanced version of Stream with the \texttt{Mem-Aware} fusion scheme to rapidly evaluate a wide variety of potential hardware configurations.

The explored design space is characterized by two primary axes: total chip area and the proportion of area allocated to memory. To traverse the first axis, the total area is varied from 12.5\% to 125\% of MARCA's total area of \SI{222}{\milli\meter\squared}. For the second axis, processing elements are substituted for additional memory capacity based on the relative compute/memory area costs reported for MARCA~\cite{li2024marca}. 
We further assume that the off-chip bandwidth scales linearly with the chip’s perimeter (commonly referred to as the \textit{beachfront} in chiplet-based architectures), which corresponds to the square root of the total area.

\begin{figure}[tb]
  \centering
  \subcaptionbox{L=1 (decode)\label{fig:contour1}}{
    \includegraphics[width=0.95\columnwidth]{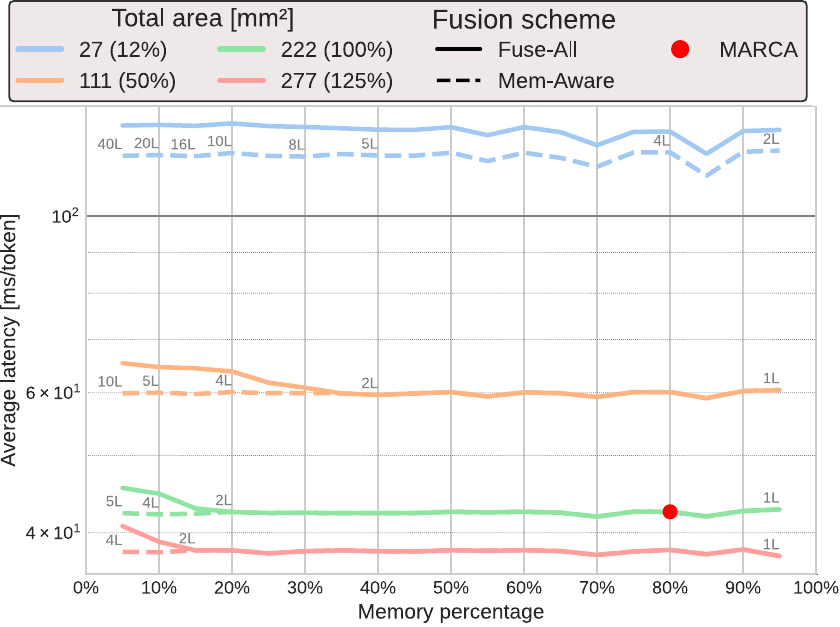}
    }
  \subcaptionbox{L=64 (prefill)\label{fig:contour64}}{%
    \includegraphics[width=0.95\columnwidth]{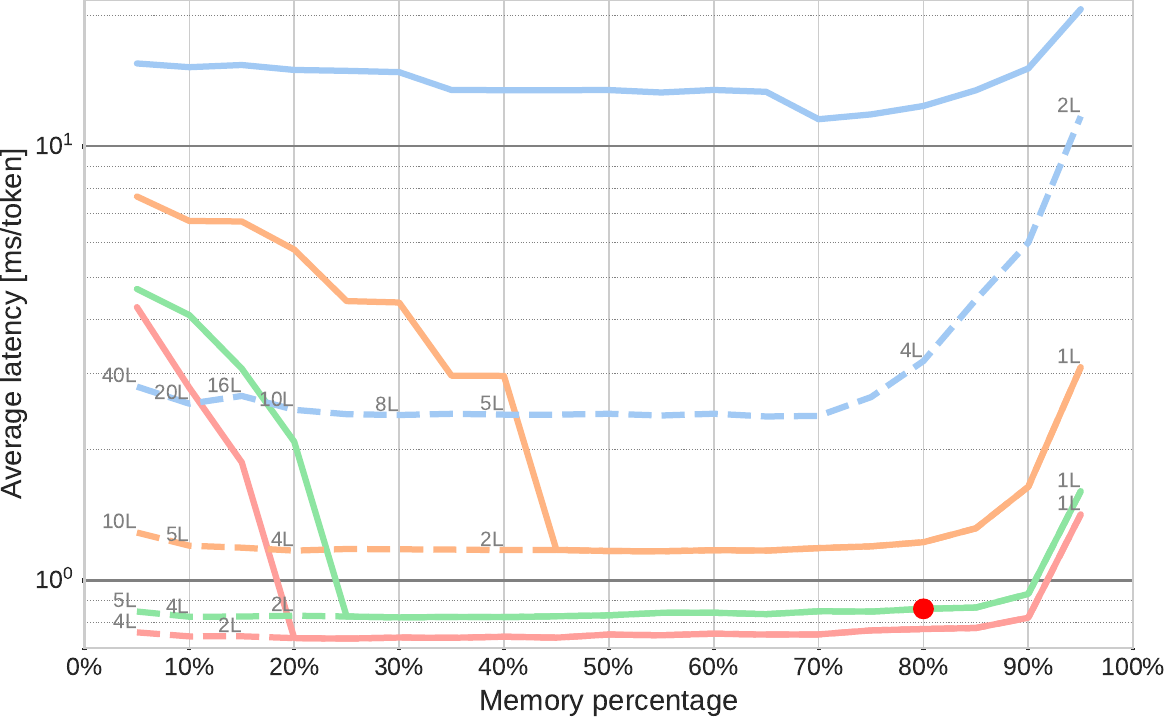}
    }
    \subcaptionbox{L=1024 (prefill)\label{fig:contour1024}}{%
    \includegraphics[width=0.95\columnwidth]{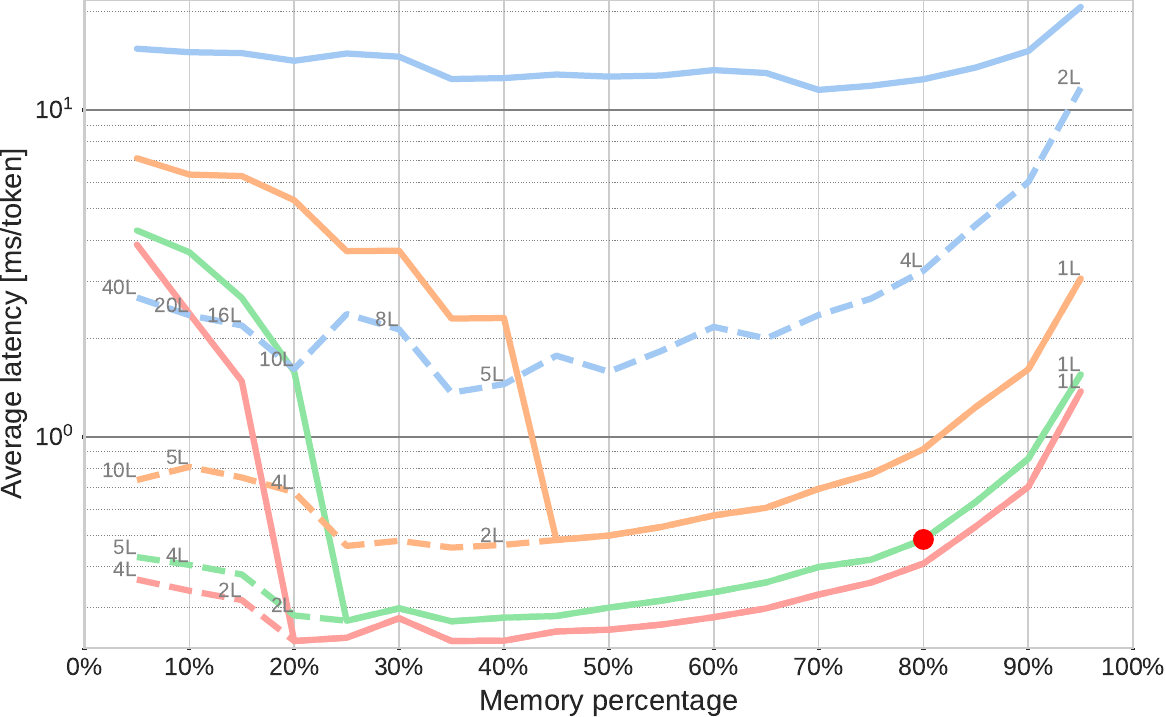}
  }
  \vspace{-3.5mm}
  \caption{Latency evaluations for different architecture design points. At the left-end side of the x-axis, the full area budget is spent on processing elements with no on-chip memory. Full lines represent fusion scheme \texttt{Fuse-All} while dashed lines represent \texttt{Mem-Aware}. Grey values indicate the total number of tiles per fused tensor for points on the dashed line. }
  \vspace{-5mm}
  \label{fig:contour}
\end{figure}

Figure \ref{fig:contour} presents the performance evaluation in the design space for the fusion schemes \texttt{Fuse-All} (full lines) and \texttt{Mem-Aware} (dashed lines), for different sequence lengths. When on-chip memory capacity is sufficiently large, the two fusion schemes exhibit identical performance, as tensors fit entirely in memory without requiring additional splits along the $D$ dimension. However, once the memory capacity falls below the requirement established in Equation~(\ref{eq:mem-size}), the performance of \texttt{Fuse-All} deteriorates.

A key observation in Figures \ref{fig:contour1} ($L\!=\!1$) and \ref{fig:contour64} ($L\!=\!64$) is the presence of a broad latency plateau along the iso-area lines, rather than a sharply defined optimal point. In these cases, redistributing chip area between memory and compute units does not improve performance, as the system is constrained by off-chip bandwidth. The performance of these short sequence lengths is memory-bound by the projection layers, meaning that increasing neither on-chip compute nor on-chip memory capacity mitigates the bottleneck.

\inlineheading{Takeaway 6} For small sequence lengths, where memory-bound projection operators limit the achievable performance, changing the hardware architecture (memory capacity or the total number of PEs) doesn't improve performance. 

For larger sequence lengths, this off-chip bandwidth limitation diminishes, as the projection weight matrix can be reused across multiple tokens, shifting the projection layers from a memory-bound to a compute-bound regime. In Figure~\ref{fig:contour1024} ($L\!=\!1024$), the set of near-optimal design points becomes significantly narrower. In particular, the MARCA architecture is positioned far from this optimal range. Under fusion scheme \texttt{Fuse-All}, the need for extensive on-chip memory is substantially reduced, allowing silicon area to be reallocated toward higher compute capacity. Specifically, an accelerator featuring $32,768$ PEs and \SI{10.5}{\mebi\byte} of SRAM would achieve a 1.78× performance improvement over MARCA. 

\inlineheading{Takeaway 7} For large sequence lengths and under the \texttt{Mem-Aware} fusion scheme, the optimal hardware architecture configuration shifts, favoring computation units over on-chip memory and leading to a significant improvement in performance. 

\cut{The \texttt{Mem-Aware} fusion scheme shifts the optimal hardware architecture for large sequence lengths, favoring more compute over on-chip memory capacity.}

%% file: 7-concls.tex
\section{Conclusion} \label{sec:concl}

SSMs have recently emerged as a promising alternative to transformers for long-sequence processing, though efficient inference with hardware accelerators remains largely unexplored. 
This work demonstrated how tiling and operator fusion can transform SSM acceleration, shifting state update computations from a memory-bound to a compute-bound regime, thereby significantly improving performance. By employing an extended version of Stream, an accelerator modeling framework, we identified key data locality bottlenecks and proposed advanced fusion and scheduling strategies to mitigate them. 

Our results show that our memory-aware fusion scheme enables significant reductions in on-chip memory capacity without sacrificing performance. In a hardware design exploration study, we identified that a fusion-aware SSM accelerator favors compute over on-chip memory to deliver increased throughput, compared to contemporary accelerator architectures. While the state-of-the-art MARCA accelerator provides a strong baseline, our findings highlight opportunities for further architectural improvements.
These insights pave the way for more efficient hardware architectures, particularly for handling larger sequence lengths in the prefill stage.
